\documentclass[iop,apl,twocolumn,showpacs,amsmath]{revtex4}
\usepackage{graphicx}

\begin{document}

\title{Ti-enhanced kinetics of hydrogen absorption and desorption on
NaAlH$_4$ surfaces}

\author{Jorge \'I\~niguez$^1$ and Taner Yildirim$^2$}

\affiliation{$^{1}$Institut de Ciencia de Materials de Barcelona
(CSIC), Campus de la UAB, 08193 Bellaterra, Spain\\ $^{2}$NIST Center
for Neutron Research, National Institute of Standards and Technology,
Gaithersburg, MD 20899, USA}

\begin{abstract}
We report a first-principles study of the energetics of hydrogen
absorption and desorption (i.e. H-vacancy formation) on pure and
Ti-doped sodium alanate (NaAlH$_4$) surfaces. We find that the Ti atom
facilitates the dissociation of H$_2$ molecules as well as the
adsorption of H atoms. In addition, the dopant makes it energetically
more favorable to creat H vacancies by saturating Al dangling
bonds. Interestingly, our results show that the Ti dopant brings close
in energy all the {\sl steps} presumably involved in the absorption
and desorption of hydrogen, thus facilitating both and enhancing the
reaction kinetics of the alanates. We also discuss the possibility of
using other light transition metals (Sc, V, and Cr) as dopants.
\end{abstract}

\pacs{81.05.Zx, 81.05.Je, 61.12.-q, 63.20.Dj} \maketitle

In recent years sodium alanate (NaAlH$_4$) has become one of the most
promising systems to achieve safe and inexpensive storage of hydrogen
aboard vehicles. The kinetics of the reversible reaction by which pure
NaAlH$_4$ releases and stores hydrogen is relatively slow. However, it
was discovered that a few percent of Ti doping increases the reaction
rates dramatically~\cite{bog97}, bringing the system close to what is
required for practical applications. In spite of extensive
investigations, the mechanism by which Ti enhances the reaction
kinetics remains largely unknown~\cite{gro02a,kiy03,gro03}. As a
matter of fact, only recently have we had convincing
experimental~\cite{gra04} and theoretical~\cite{lov05,ini05} evidence
that the Ti dopants are located on the surface, as opposed to the
bulk, of the system. First-principles simulations by several groups
also predict that Ti dopants cause the breaking of multiple Al--H
bonds~\cite{ini05,ini04}, and that bulk Ti dopants~\cite{moy05a} and
Na vacancies~\cite{moy05b} facilitate the creation of H
vacancies. Such studies suggest mechanisms by which Ti can improve the
H desorption kinetics of the alantes, but are far from being complete
or conclusive (specially those restricted to the bulk of the system).

In this work we present a first-principles study of the energetics of
hydrogen absorption and desorption (i.e. vacancy formation) on pure
and Ti-doped NaAlH$_4$ sufaces. We find that the Ti dopant facilitates
{\sl both} the absorption and the desorption, thus enhancing the
cycling kinetics of hydrogen.


The calculations were performed within the generalized gradient
approximation~\cite{per96} to density functional theory~\cite{dft} as
implemented in the code SIESTA~\cite{sol02}. We used a localized basis
set including double-$\zeta$ and polarization orbitals, and
Troullier-Martins pseudopotentials~\cite{tro93}. We tested the
convergence of our calculations with respect to the k-point and real
space meshes: we typically considered a 4500~\AA$^3$ simulation
supercell containing about 162 atoms, for which we used a
2$\times$2$\times$1 k-point grid and a 150~Ry cutoff for the real
space mesh. The accuracy of our SIESTA resutls had been previously
tested against more accurate {\sl ab initio} calculations using a
plane-wave basis set~\cite{ini05}.


Tetragonal NaAlH$_4$ presents two natural terminations related to the
(100) and (001) directions, respectively. The phenomenology associated
with the Ti-doping of these two surfaces is known to be essentially
the same~\cite{ini05}; thus, for this work we considered the
001-surface, which can be treated using a relatively small
supercell. Our slab-type supercell can be regarded as composed of
3$\times$3$\times$3 NaAlH$_4$ groups, i.e., 162 atoms. We imposed that
the atoms in the deepest layer be fixed at the bulk atomic positions,
and allowed about 13~\AA\ of empty space between slabs. The bulk
structural parameters ($a=b=$5.01~\AA\ and $c=$11.12~\AA) were
obtained from first-principles~\cite{ini05}. Structural relaxations
were considered to be converged for residual force components smaller
than 0.02~eV/\AA. We calculated absorption and vacancy-formation
energies as the difference between the energies of products and
reactants; thus, a negative value signals an energetically favorable
reaction. To study, for example, the deposition of H atoms, we
considered several starting configurations, performed regular
structural relaxations, and took the lowest-energy result. Ideally one
would use, for example, simulated annealing techniques to identify the
true lowest-energy structures, but such an approach is too
computationally costly.

The question of whether the Ti dopant substitutes for Na or Al is
somehow controversial, as it critaically depends on the choice of
energies that are used as a reference to compute the relative
stability of the various doping models~\cite{moy05b}. The Ti doping of
the alanates is usually done by ball-milling, which is an
out-of-equilibrium process that involves individual Ti, Na, and Al
atoms, as well as H$_2$ molecules, rather than the corresponding
crystalline phases. Hence, we prefer using atomic energies as
reference, which leads to the conclusion that Ti substitutes for
Na~\cite{ini05,ini04}. However, if the energies of crystalline Ti, Na,
and Al are used, the result is that Ti substitutes for
Al~\cite{lov05}. Being aware of this issue, in this work we have
considered these two Ti-doping models, i.e. Ti substituting for Na
(Ti@Na) and Ti substituting for Al (Ti@Al), and found that both lead
to the same qualitative results. Note also that our present discussion
is based on how the pure and doped surfaces differ regarding the
energetics of hydrogen absorption and vacancy formation. Our
conclusions are thus independent of the chosen energy reference, which
is the same for both pure and doped surfaces.


{\sl Hydrogen absorption.--} We first studied the possibility that the
alanate surface, pure and Ti-doped, might absorb extra hydrogens. The
underlying assumption is that a favorable absorption energetics will
facilitate the NaAlH$_4$-formation reaction, which occurs under H$_2$
pressure. We considered the direct absorption of an H$_2$ molecule as
well as the absorption of individual H atoms that we deposited on the
surface one at a time. Table~\ref{tab1} summarizes our results.

\begin{table}[t]
\caption{Absorption energies (in eV) of molecular H$_2$ and various
  $n$H cases (where the H atoms are deposited on the surface one at a
  time) of pure and Ti-doped NaAlH$_4$ surfaces. Two doping models are
  considered, in which Ti substitutes for Na (Ti@Na) and Al (Ti@Al),
  respectively. We use as reference the energy of the surfaces with no
  extra hydrogen, and the energy of the H$_2$ molecule. See text for
  the explanation of the Ti@Na+H$_2$ results.}
\vskip 2mm
\begin{tabular*}{1.0\columnwidth}{@{\extracolsep{\fill}}lcrrrr}
\hline\hline
Surf. type    & +H$_2$ & +H  & +2H & +3H & +4H \\
\hline
pure       & $-$0.07 & 1.81 &      &      &      \\
Ti@Na      & $-$0.03/$-$0.23 & $-$0.49 & $-$0.23 & $-$0.67 & $-$0.38 \\
Ti@Al      & $-$0.41 & $-$0.61 & $-$0.41 & $-$0.60 &         \\
\hline\hline
\end{tabular*}
\label{tab1}
\end{table}

\begin{figure}[t]
\includegraphics[width=8cm]{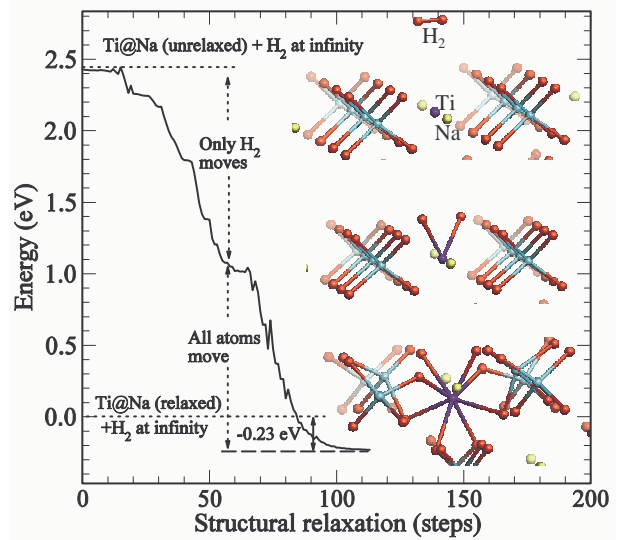}
\caption{(color online) Energy {\sl versus} structural relaxation step
for the H$_2$ dissociation and absorption by the Ti@Na surface. The
starting point is the {\sl unrelaxed} Ti@Na surface with an H$_2$
molecule close to it, which is taken as the zero of energy. In the
first part of the relaxation, only the H$_2$ molecule is allowed to
move; then, all the atomic positions are relaxed. Given for reference
is the result for the relaxed Ti@Na plus an H$_2$ located at
infinity. The insets show the various steps of the absorption
process.}
\label{fig:h2absorption}
\end{figure}

We found that the surface is able to bind an H$_2$ molecule in all the
cases considered. The molecule dissociates in the Ti@Al case, forming
an AlH$_6$ group, which renders a significant energy gain of 0.41~eV.
In the case of the pure surface, the H$_2$ molecule does not
dissociate and the binding energy is significantly smaller
(0.07~eV). Finally, the Ti@Na case turns out to be more complicated:
When the H$_2$ molecule is deposited on the relaxed Ti@Na surface, it
does not {\sl see} the Ti atom and binds weakly without dissociating
(the calculated binding energy is 0.03~eV). However, if the H$_2$
molecule is deposited on the {\sl unrelaxed} Ti@Na surface, it
dissociates and binds with an energy of 0.23~eV. It is useful to think
of this absorption as the two-step process depicted in
Fig.~\ref{fig:h2absorption}. In the first step the molecule breaks and
the H atoms bind to the Ti on the alanate surface. The corresponding
energy gain is about 1.4~eV, which is very close to what is obtained
for H$_2$ breaking and absorption on carbon nanostructures coated with
Ti~\cite{yil05}. In the second step the whole surface relaxes and the
Ti dopant adopts a final configuration with 8 neighboring H atoms (see
botom inset of Fig.~\ref{fig:h2absorption}). In conclusion, our
results for both Ti@Al and Ti@Na clearly show that the Ti dopand
facilitates H$_2$ dissociation and absorption.

Next we studied the deposition of individual H atoms one at a time. We
found that the pure surface is unable to absorb an individual H atom
(note the positive absorption energy of 1.8~eV in Table~\ref{tab1});
in fact, the deposited H destabilizes an AlH$_4$ group and forms an
H$_2$ molecule. In contrast, the doped surfaces are able to absorb the
first H atom with significant energy gain (about 0.5~eV); the H atom
locates in the vicinity of the Ti dopant, at about 1.9~\AA, forming a
TiH$_5$ group in the case of Ti@Al. Moreover, further hydrogens can be
absorbed by the doped surfaces. For Ti@Na, the +3H situation seems to
be particularly stable; for +4H an H$_2$ molecule forms and stays
bound to the surface, but not properly absorbed by it. For Ti@Al, we
found that a H$_2$ molecule forms already in the +3H case. Thus, our
results indicate that a Ti dopant can take a maximum of 3 or 4 extra
hydrogens.

In order to enhance the NaAlH$_4$-formation kinetics, the absorbed
hydrogens should be {\sl mobile} on the system surface. To check this,
we performed computer experiments as the following one: we start from
the relaxed Ti@Na+2H system, create a H vacancy and thus a defective
AlH$_3$ group, relax the system (H migration should occur to recover
an AlH$_4$ group), and compare the obtained structure and energy with
the {\sl assumed} lowest-energy solution Ti@Na+H. We observed two
general trends. (1) Large structural distortions and atom migrations
occur. Most relaxed structures differ significantly from the
lowest-energy solution and are slightly above in energy (by about
0.1~eV, typically). (2) Hydrogen migration occurs more easily in
H-rich surfaces, i.e., in the +3H and +4H cases of
Table~\ref{tab1}. The difficulties to reach the reference,
lowest-energy structure reflect the fact that we are dealing with a
complicated multi-minima energy landscape. On the other hand, the
significant atomic mobility observed suggest that, at room
temperature, the H atoms initially bound to the Ti dopant will indeed
migrate and take part in the NaAlH$_4$-formation reaction, thus
increasing its rate.

\begin{table}[t]
\caption{Desorption energies (in eV) of hydrogen and other atom groups
  for pure and two Ti-doped NaAlH$_4$ surfaces. We denote by +H$^v$
  the surface with one H vacancy, etc. We use as reference the energy
  of the original pure and Ti-doped surfaces, the energy of the H$_2$
  molecule, and the atomic energies of Na and Al.}
\vskip 2mm
\begin{tabular*}{0.80\columnwidth}{@{\extracolsep{\fill}}lcccc}
\hline\hline
Surf. type    & +H$^v$ & +Na$^v$  & +(NaH)$^v$ & +(AlH$_4$)$^v$ \\
\hline
pure       &    1.98  &  7.96  &  5.61  &  7.51  \\ 
Ti@Na      &    0.41  &  4.89  &  5.41  &  5.70  \\
Ti@Al      &    0.30  &  5.20  &  4.50  &  5.65  \\
\hline\hline
\end{tabular*}
\label{tab2}
\end{table}

{\sl Hydrogen vacancy formation.--} Next we studied the ability of the
Ti dopant to facilitate the formation of hydrogen vacancies and,
further, destabilize the surface. We calculated the formation energies
of various types of vacancies: H, Na, and atom groups like NaH and
AlH$_4$. (We denote the H vacancy by H$^v$, etc.) Table~\ref{tab2}
summarizes our results.

We first note that all the vacancy-formation energies are positive;
thus, as expected, we find that the surface, either pure or doped, is
stable and thermal activation energy is required to decompose it. The
second observation is that the pure surface is significantly more
stable than the Ti-doped surfaces. For example, the formation energy
of H vacancies is about 2.0~eV for the pure surface, while it is below
0.5~eV when the Ti dopant is present. The same trend is observed in
the formation energy of the other studied {\sl defects}, which clearly
indicates that the Ti-doped surface will decompose at temperatures
significantly lower than those required in the case of a pure system.

\begin{figure}[t]
\includegraphics[width=\columnwidth]{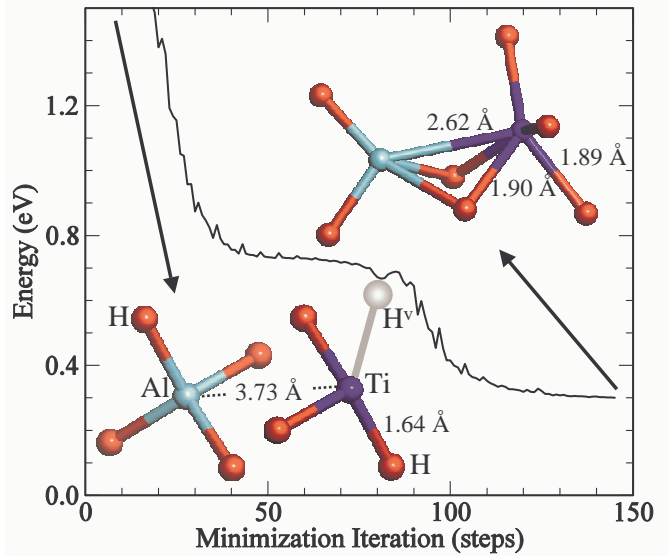}
\caption{(color online) Energy {\sl versus} structural relaxation step
for the relaxation of a Ti@Al surface in which we create a hydrogen
vacancy neighboring the Ti atom (we denote this case by
Ti@Al+H$^v$). The formation of an Al--Ti bond in the final
configuration results in a reduced H$^v$ formation energy. The insets
show the initial and final configurations of the atoms mainly involved
in the structural relaxation.}
\label{fig:passivation}
\end{figure}

The Ti dopant is able to reduce the defect formation energies thanks
to its chemical versatility. In the case of {\sl charged} defects, the
dopant adquires extra charge (e.g. in the Ti@Na+(AlH$_4$)$^v$ case) or
enables effective passivation of dangling bonds. The latter is best
illustrated by the Ti@Al+H$^v$ case depicted in
Fig.~\ref{fig:passivation}: The relaxed configuration presents the Ti
and a neighboring Al atom within bonding distance (2.62~\AA) and
sharing two hydrogen atoms. This relaxation results in a significant
energy reduction and thus facilitate the H$^v$ formation. In the case
of {\sl neutral} defects (e.g. (NaH)$^v$) the Ti dopant does not seem
to play such a relevant role and, thus, the pure and doped surfaces
present relatively similar vacancy-formation energies.

Note that we considered explicitly the case of Na-vacancy mediated H
desorption, which has recently been claimed to be specially important
on the basis of bulk NaAlH$_4$ calculations~\cite{moy05b}. Indeed, we
found that Na-defective pure and Ti@Al surfaces gain energy by
releasing a hydrogen atom (about 2.3 and 0.7~eV, respectively), while
there is an energy cost of 0.5~eV in the Ti@Na case. On the other
hand, the occurrence of the Na vacancy is much more likely when the Ti
dopant is present, for both Ti@Al and Ti@Na, which renders the net
(NaH)$^v$ formation energies in Table~\ref{tab2}. As already
mentioned, since (NaH)$^v$ is not a charged defect, there is not a
very big difference between the results for pure and doped
surfaces. Our calculations indicate that the Ti dopand favors the
Na$^v$-mediated mechanism for H desorption slightly in the Ti@Na case
and more significantly in the Ti@Al case.

Our calculations render a conclusion that may seem paradoxical at
first, namely, that the Ti-dopant favors {\sl both} the absorption and
desorption of hydrogen on NaAlH$_4$ surfaces, thus enhancing the
kinetics of the H-charge and discharge reactions of the system. The
explanation to this paradox is that, as a result of its chemical
versatility, the Ti dopant brings close in energy all the {\sl steps}
presumably involved in these reactions, with the corresponding
reduction in activation energies and, thus, increased reaction
rates. For this picture to hold, we need to assume that the Ti dopants
remain on the surface of the system during the reactions, which seems
plausible given the large ionic mobility involved in these processes
and the fact that, as shown in Ref.~\cite{ini05}, it is energetically
favorable for the dopant to lie on the surface of the system.


Finally, we studied the possibility of using, instead of Ti, other
light transition metals as dopants. More precisely, we tested the
performance of Sc, V, and Cr by calculating the energies of absorption
and desorption of a single H atom by a Na-substituted
surface. Regarding absorption, only the Sc dopant favors it, with a
negative absorption energy of $-$0.27~eV; we get 0.26 and 0.20~eV for
V and Cr, respectively. We also found that Sc causes the H$_2$
molecule to dissociate provided H$_2$ is deposited on the unrelaxed
Sc@Na structure. As for the formation of vacancies, the lowest energy
is obtained for Cr with 0.30~eV; V gives 0.75~eV and Sc renders the
most stable surface, with a formation energy of 1.13~eV. Hence, a
comparison with the pure surface results in Tables \ref{tab1} and
\ref{tab2} indicates that Sc, V, and Cr will facilitate both hydrogen
absorption and desorption. At the same time, our results clearly show
that Ti performs better in all aspects, thus being the dopant of
choice among light transition metal atoms.


We acknowledge financial support from the Spanish Ministry of Science
and Education through the ``Ram\'on y Cajal'' program (JI) and project
BFM2003-03372-C03-03, the Catalan Regional Government through project
2005SGR683, FAME-NoE, and the U.S. DoE under BES grant
DE-FG02-98ER45701. Use was made of the facilities provided by the
CESGA Supercomputing Center.


\begin{thebibliography}{99}

\bibitem{bog97} B. Bogdanovic and M. Schwickardi, J. Alloys Comp. {\bf 253},
  1 (1997).

\bibitem{gro02a} K.J. Gross, G.J. Thomas, and C.M. Jensen, J. Alloys
  Compd. {\bf 330-332}, 683 (2002).
  
\bibitem{kiy03} T. Kiyobayashi, S.S. Srinivasan, D. Sun, and
C.M. Jensen, J. Phys. Chem. A {\bf 107}, 7671 (2003).

\bibitem{gro03} K.J. Gross, E.H. Majzoub, and S.W. Spangler, J. Alloys
Compd. {\bf 356}, 423 (2003).

\bibitem{gra04} J. Graetz, J.J. Reilly, J. Johnson, A.Y. Ignatov, and
T.A. Tyson, Appl. Phys. Lett. {\bf 85}, 500 (2004).

\bibitem{lov05} O.M. Lovvik and S. Opalka, Phys. Rev. B {\bf 71},
  054103 (2005).

\bibitem{ini05} J. \'I\~niguez and T. Yildirim, Appl. Phys. Lett. {\bf
  86}, 103109 (2005).

\bibitem{ini04} J. \'I\~niguez, T. Yildirim, T.J. Udovic, M. Sulic,
and C.M. Jensen, Phys. Rev. B {\bf 70}, 060101(R) (2004).

\bibitem{moy05a} C.M. Ara\'ujo, R. Ahuja, and J.M. Osorio Guill\'en,
  Appl. Phys. Lett. {\bf 86}, 251913 (2005).

\bibitem{moy05b} C.M. Ara\'ujo, S. Li, R. Ahuja, and P. Jena,
  Phys. Rev. B {\bf 72}, 165101 (2005).

\bibitem{per96} J.P. Perdew, K. Burke, and M. Ernzerhof,
Phys. Rev. Lett. {\bf 77}, 3865 (1996).

\bibitem{dft} P. Hohenberg and W. Kohn, Phys. Rev. {\bf B864}, 136
  (1964); W. Kohn and L.J. Sham, Phys. Rev. {\bf A1133}, 140 (1965).

\bibitem{sol02} J.M. Soler {\sl et al.}, J. Phys.: Cond. Matt. {\bf
14}, 2745 (2002).

\bibitem{tro93} N. Troullier and J.L. Martins, Phys. Rev. B {\bf 43}, 1993
(1991).

\bibitem{yil05} T. Yildirim and S. Ciraci, Phys. Rev. Lett {\bf 94},
  175501 (2005); T. Yildirim, J. \'I\~niguez, and S. Ciraci,
  Phys. Rev. B {\bf 72}, 153403 (2005).

\end{thebibliography}
\end{document}